\documentclass[journal]{IEEEtran}
\usepackage{cite}
\usepackage[pdftex]{graphicx}
\DeclareGraphicsExtensions{.pdf,.jpeg,.png}
\graphicspath{{./fig/}}
\usepackage{amsmath}
\usepackage{autobreak}
\usepackage{amsthm}
\usepackage{amssymb}
\usepackage{bbm}
\usepackage[ruled]{algorithm2e}
\usepackage{array}
\usepackage{cleveref}
\ifCLASSOPTIONcompsoc
  \usepackage[caption=false,font=normalsize,labelfont=sf,textfont=sf]{subfig}
\else
  \usepackage[caption=false,font=footnotesize]{subfig}
\fi
\usepackage{threeparttable}
\usepackage{tabularx}
\usepackage{multirow}
\usepackage{multicol}
\usepackage{booktabs}
\usepackage{url}
\usepackage[x11names]{xcolor}
\usepackage{soul}
\soulregister\cite7
\soulregister\cref7
\soulregister\Cref7

\usepackage{tikz}
\usetikzlibrary{calc}
\usepackage{lipsum}

\makeatletter 
\newcount\SOUL@minus
\makeatother  

\begin{document}
\title{Online Multi-agent Reinforcement Learning for Decentralized Inverter-based Volt-VAR Control}
\author{Haotian~Liu, ~\IEEEmembership{Graduate~Student~Member,~IEEE},
  Wenchuan~Wu,~\IEEEmembership{Fellow,~IEEE}
  \thanks{This work was supported by the National Key R\&D Program of China under Grant 2020YFF0305800.}
  \thanks{H. Liu and W. Wu (Corresponding Author)  are with the State Key Laboratory of Power Systems, Department of Electrical Engineering, Tsinghua University, Beijing 100084, China (email:lht18@mails.tsinghua.edu.cn, wuwench@tsinghua.edu.cn).}
}
\markboth{IEEE TRANSACTIONS ON SMART GRID,~Vol.~xx, No.~x, February~2021}%
{Shell \MakeLowercase{\textit{et al.}}: Bare Demo of IEEEtran.cls for IEEE Journals}
\maketitle

\begin{abstract}
  The distributed Volt/Var control (VVC) methods have been widely studied for active distribution networks(ADNs),
  which is based on perfect model and real-time P2P communication.
  However, the model is always incomplete with significant parameter errors and such P2P communication system is hard to maintain.
  In this paper, we propose an online multi-agent reinforcement learning and decentralized control framework (OLDC) for VVC. In this framework,
  the VVC problem is formulated as a constrained Markov game and we propose a novel multi-agent constrained soft actor-critic (MACSAC) reinforcement learning algorithm.
  MACSAC is used to train the control agents online, so the accurate ADN model is no longer needed.
  Then, the trained agents can realize decentralized optimization using local measurements without real-time P2P communication.
  The OLDC with MACSAC has shown extraordinary flexibility, efficiency and robustness to various computing and communication conditions.
  Numerical simulations on IEEE test cases not only demonstrate that the proposed MACSAC outperforms the state-of-art learning algorithms,
  but also support the superiority of our OLDC framework in the online application.
\end{abstract}

\IEEEpeerreviewmaketitle

\section{Introduction}
\IEEEPARstart{V}{oltage} violation problems and high network losses are becoming increasingly severe in active distribution networks (ADN) with high penetration level of distributed generation (DG)\cite{6336354,8960272}.
As an important solution, Volt-VAR control (VVC) has been successfully integrated into distribution management systems to optimize the voltage profile and reduce network losses.
Since most DGs are inverter-based energy resources (IB-ERs), they are able and required to provide fast Volt/VAR support using their free capacity.

Conventionally, VVC is described as a nonlinear programming problem to generate a set of optimal strategies for voltage regulation devices and reactive power resources.
Plenty of literatures solve VVC problems using centralized optimization methods such as interior point methods \cite{4808225} and evolutionary algorithms \cite{6336354}.
Despite the wide application of centralized VVC, they suffered from the single-point failure and heavy computation \& communication burdens.
Also, as for the increasingly huge amount of IB-ERs, centralized VVC is also limited with communication-dependent time-delay issues.

Therefore, distributed VVC methods have been proposed to exploit the distributed nature of the ADN.
Distributed methods utilizes local measurements with P2P communication with neighbors to realize fast control.
Previous papers mainly adapt distributed optimization algorithms, such as quasi real-time reactive optimization \cite{7470528}, alternating direction method of multipliers (ADMM) \cite{7926415,7042735} and accelerate ADMM \cite{8960272}.
However, these P2P communication system is hard to maintain in real practice.
There are also some decentralized methods \cite{7361761,7500071} to realize quasi-optimal control, which are based on improved droop control strategies and only local measurements are used for each controller.

Till now, most VVC algorithms depend on the accurate ADN models to achieve desirable performance.
It is impractical and expensive for regional power utilities to maintain such reliable models, especially in a distribution system with increasing complexity and numerous buses \cite{6579907,8944292}.
Recently, the effectiveness of (deep) reinforcement learning (RL) based approaches have been verified to cope with the incomplete model challenges in energy trading \cite{8661902}, emergency control \cite{8787888}, load frequency control \cite{8534442}, and voltage regulation \cite{8944292,8873679}.

In order to apply RL algorithms in a distributed or decentralized manner, multi-agent RL has been studied in inspiring attempts \cite{5589633,6392462,8331897,liuDistributedReinforcementLearning2018,doi:10.1177/0037549710367904,8981922,9076841}.
\cite{liuDistributedReinforcementLearning2018} proposes a distributed reinforcement learning based secondary control of DC microgrids based on pinning consensus.
\cite{8981922} develops a decentralized cooperative control strategy for multiple energy storage systems based on Q-learning and the value decomposition network (VDN) from \cite{sunehag2017valuedecomposition}.
\cite{9076841} develops a multi-agent autonomous voltage control method based on the state-of-art algorithm multi-agent deep deterministic policy gradient (MADDPG) proposed in \cite{NIPS2017_7217}. Besides, data-driven decentralized control is also studied in \cite{baghaeeUnbalancedHarmonicPower2018,baghaeePowerCalculationUsing2016,baghaeeDecentralizedRobustMixed2018} for dynamic voltage controls to reduce voltage harmonics and improve power quality.

However, the existing control methods either a) implement training of agents in the offline stage based on a simulation model and execute them online without training, which sacrifice the model-free feature of multi-agent RL,
or b) synchronously learn the agents online with heavy communication burdens.
As for VVC with numerous high-speed IB-ERs, a novel online multi-agent RL framework that performs online learning without heavy communication and local computation burdens is urgently desired.

Moreover, to realize such framework, there are several critical technical challenges:
\begin{enumerate}
  \item The deterministic policies of Q-learning and DDPG algorithms lead to extreme brittleness and notorious hyperparameter sensitivity \cite{haarnoja2018sacapplications}, which limit the online application.
  \item The power system operational constraints are not modelled explicitly in the existing multi-agent RL based methods, which is a critical issue in VVC.
  \item Online exploration of the data-driven algorithms could lead to deterioration on the performance of VVC. Such exploration and exploitation issue is especially serious in ADN with high speed IB-ERs.
\end{enumerate}

\begin{figure}[htbp]
  \centering
  \includegraphics[width=0.95\linewidth]{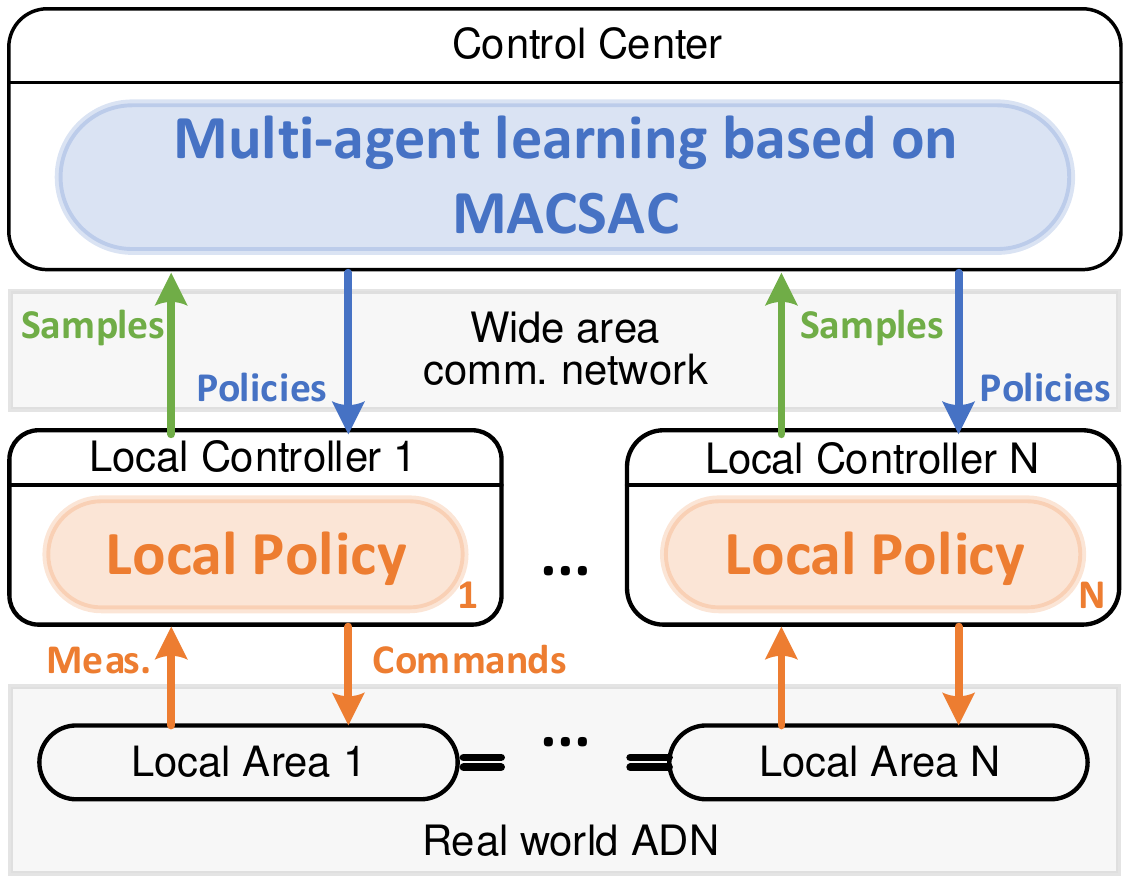}
  \caption{Overall structure of the proposed online multi-agent learning and decentralized control framework.}
  \label{fig:proposed}
\end{figure}

In this paper, we propose an \textbf{O}nline multi-agent reinforcement \textbf{L}earning and \textbf{D}ecentralized \textbf{C}ontrol framework (OLDC) for VVC as shown in \cref{fig:proposed}.
Moreover, to improve the stability and efficiency of VVC, we propose a novel multi-agent RL algorithm called Multi-Agent Constrained Soft Actor-Critic (MACSAC) inspired by previous works \cite{haarnoja2018sacapplications,10.5555/3305381.3305384,8944292,NIPS2017_7217}.

As shown in \cref{fig:proposed}, coordinated multi-agent learning based on MACSAC is conducted in the control center, and the latest trained polices are sent to controllers to carry out local control.
With the asynchronous learning, sampling and control processes, this solution can realize safe and fast model free optimization for VVC in ADNs.
The unique contributions of this article are summarized as follows.
\begin{enumerate}
  \item Compared to the existing algorithms like MADDPG \cite{NIPS2017_7217}, our proposed MACSAC significantly improves the stability and efficiency of the training and application processes.
  Instead of using deterministic policies, MACSAC utilizes stochastic policies with maximum entropy regularization following \cite{haarnoja2018sacapplications}, which prevents optimization failure and ameliorates training robustness.
  MACSAC also explicitly model voltage constraints following \cite{8944292} instead of treat it as a penalty, which can significantly improve voltage security level.
  \item In order to synergistically combine online multi-agent RL and decentralized control, a novel OLDC framework with detailed timing design is proposed in this paper.
  The proposed VVC with OLDC can both learn the control experiences continuously to meet the incomplete model challenge, and make decision locally to realize fast control.
  Also, OLDC can be extended to apply in other multi-agent power system controls, and is capable with future off-policy multi-agent RL algorithms.
  \item With the off-policy nature of MACSAC, our OLDC provides a promising method for balancing exploration and exploitation in RL-based algorithms.
  The safety and operation efficiency is dramatically enhanced by saving the cost of redundant exploration online.
\end{enumerate}

The remainder of this article is organized as follows. \Cref{sec:prelim} formulates the the VVC problem in ADNs as a constrained multi-agent Markov game, and also briefly introduces RL and the multi-agent actor-critic framework. Then, the detailed introduction to the proposed MACSAC and OLDC are presented in \Cref{sec:methods}. In \Cref{sec:numerical} the results of our numerical study are shown and analyzed. Finally, \Cref{sec:conclusion} concludes this article.

\section{Preliminaries}
In this section, we firstly introduce the VVC problem in this paper. 
Then, the settings of Markov games and RL in this paper is explained.
In the last subsection, we introduce preliminaries of actor-critic and multi-agent actor-critic methods.

\label{sec:prelim}
\subsection{VVC Problem Formulation}
An ADN is divided into $N$ nature control areas with local measurements and control agent.
It can be depicted by an undirected graph $\Pi(\mathcal{N}, \mathcal{E})$ with the collection of all nodes $\mathcal{N}=\bigcup_{i\in[1,N]} \mathcal{N}_i$, the collection of each area $i$'s nodes $\mathcal{N}_i$, and the collection of all branches $\mathcal{E}$.
Since it is common for the ADN in the real world to equip only with single-phase steady-state measurements, the VVC problem is formulated on balanced networks for real-time steady-state dispatch in this paper.
Since the inner details of the model are not required and only the input and output data are necessary, such model can be easily extended to unbalanced multi-phase networks. Such extension is validated in \cref{sub:3phase} with a three phase unbalanced distribution network.

While we consider the steady-state voltage control, the power flow equations are employed as shown in \cref{eq:pf}, where $P_{ij},Q_{ij}$ is the active and reactive power flow from node $i$ to $j$, $V_i$ is the voltage at node $i$ and $G_{ij}+jB_{ij}$ is the admittance of branch ${ij}$, and $G_{sh,i} + j B_{sh, i}$ is the shunt admittance of node $i$.
\begin{equation}
  \label{eq:pf}
  \begin{split}
    P_{ij} &= G_{ij}V_i^2 - G_{ij}V_i V_j \cos{\theta_{ij}} - B_{ij}V_i V_j \sin{\theta_{ij}}, \forall ij \in \mathcal{E} \\
    Q_{ij} &= -B_{ij}V_i^2 + B_{ij}V_i V_j \cos{\theta_{ij}} - G_{ij}V_i V_j \sin{\theta_{ij}}, \forall ij \in \mathcal{E} \\
    \theta_{ij} &= \theta_i-\theta_j,\forall ij \in \mathcal{E}
  \end{split}
\end{equation}

The $k$th area is equipped with $n_\text{IBk}$ IB-ERs and $n_\text{CDk}$ compensation devices such as static Var compensators (SVC).
Without loss of generality, we assume that the IB-ERs and compensation devices are installed on different nodes in $\mathcal{N}_k$.
Accordingly, the collection of the nodes equipped with IB-ERs and compensation devices are noted as $\mathcal{N}_\text{IBk}$ and $\mathcal{N}_\text{CDk}$.

Since $\mathcal{N}_\text{IBk}\cap \mathcal{N}_\text{CDk} = \emptyset$, the power injections at each nodes can be determined via \cref{eq:injection}.
\begin{equation}
  \label{eq:injection}
  \begin{aligned}
    G_{sh,i} V_i^2 + \sum_{ij \in \mathcal{E}}P_{ij} &= \begin{cases}
      -P_{Dj}, j\in \mathcal{N} \backslash \mathcal{N}_\text{IBk} \\
      P_{Gj} - P_{Dj}, j\in \mathcal{N}_\text{IBk}
    \end{cases} \\
    -B_{sh,i} V_i^2 + \sum_{ij \in \mathcal{E}}Q_{ij}  &= \begin{cases}
      -Q_{Dj}, j\in \mathcal{N} \backslash \{\mathcal{N}_\text{IBk} \cup \mathcal{N}_\text{CDk}\} \\
      Q_{Gj} - Q_{Dj}, j\in \mathcal{N}_\text{IBk} \\
      Q_{Cj} - Q_{Dj}, j\in \mathcal{N}_\text{CDk}
    \end{cases}
  \end{aligned}
\end{equation}
where $Q_{Cj}$ is the output of reactive compensator at node $j$;
$P_{Gj},Q_{Gj}$ are the active and reactive power output of DG at node $j$;
$P_{Dj}, Q_{Dj}$ are the active and reactive power of the load at node $j$.

The IB-ERs are typically designed with redundant rated capacity for safety reasons and operate under maximum power point tracking (MPPT) mode. Hence, the controllable range of the reactive power of IB-ERs can be determined by the rated capacity $S_{Gi}$ and current active power output $P_{Gi}$. The reactive power range of controllable devices is $|Q_{Gi}| \leq \sqrt{S_{Gi}^2-P_{Gi}^2}$ and $\underline{Q_{Ci}} \leq Q_{Ci} \leq \overline{Q_{Ci}}$.

\subsection{Markov Games and Reinforcement Learning}
\label{sub:mgrl}
In order to formalize sequential multi-agent decision processes, we consider an extension of the Markov decision processes (MDP) called constrained Markov Games (CMG), which can be seen as a constrained version of Markov games (MG) \cite{LITTMAN1994157}.
In a MG, multiple agents can interact with a common environment locally.
A MG for $N$ agents is defined by a tuple $(\mathcal{S},[\mathcal{O}_i]_N,[\mathcal{A}_i]_N,\rho,[R_i]_N,\gamma)$.
The set of states $\mathcal{S}$ describes all possible states of the common environment.
The sets of local observations $\mathcal{O}_1, \dots, \mathcal{O}_N$ and actions $\mathcal{A}_1, \dots, \mathcal{A}_N$ are the local observations and actions for each agent.$R_i$ is the $i$th reward function defined as $\mathcal{S}\times\mathcal{A}_i\mapsto\mathbb{R}$.

In each time step $t$, each agent $i$ firstly observes the environment as $o_{i,t}\in \mathcal{O}_i$;
then, chooses its action $a_{i,t}$ using a stochastic policy defined as a probability density function $\pi_i:\mathcal{O}_i\times \mathcal{A}_i \mapsto [0,\infty)$, i.e., $a_{i,t}\sim\pi_i(\cdot\left|o_{i,t}\right.)$.
The actions taken at $t$ lead to the next state according to an unknown state transition probability $\rho: \mathcal{S}\times\mathcal{A}_1\times \dots \times\mathcal{A}_N\times\mathcal{S}\rightarrow [0,\infty)$.
After the transition, each agent $i$ obtains an reward $r_{i,t}$ by the corresponding reward function $R_i(s_{i,t},a_{i,t})$ and receives the next observation $o_{i,t+1}$.
The goal of each agent is to maximize its own total expected discounted return $J_i=\mathbb{E}\left[\sum_{t=0}^T \gamma^t r_{i,t}\right]$, where $\gamma$ is a discount factor and $T$ is the time horizon.
Note $s_0$ as the initial state, $\pi$ as all policies, $o_t$ as all local observations at $t$, $a_t$ as all actions at $t$ for convenience.

In the power system control domain, it is important for RL agents to keep safe exploration.
A natural way to incorporate safety is to formulate constraints into the RL problem.
Following the constrained MDP (CMDP) given by \cite{cmdp}, CMG is formulated as an constrained extension of MG, where each agent $i$ must satisfy its own constraints on expectations of auxiliary costs.
An extra group of auxiliary cost functions $R^c_1, \dots, R^c_N$ defined as $R^c_i:\mathcal{S}\times\mathcal{A}_i\mapsto\mathbb{R}$ is inserted into the tuple of MG.
At time step $t$, the cost is defined as $r^c_{i,t}=R^c_i(s_t, a_{i,t})$ where $s_t\in\mathcal{S}$.
The constraints are expressed as $J_i^c=\mathbb{E}\left[\sum_{t=0}^T \gamma^t r^c_{i,t}\right]\leq \overline{J}^c_i$.

Under the settings of CMG, the task of the RL algorithms, or multi-agent RL algorithms explicitly, is to learn an optimal policy $\pi_i^*$ for each agent $i$ to maximize $J_i$, i.e.,
\begin{equation}
  \pi_i^*(a_{i,t}\left|o_{i,t}\right.) = \arg\max_{\pi_i} J_i(\pi_i) \quad s.t.
\quad J_i^c \leq \overline{J}^c_i,
\end{equation}
with sequential decisions data and without knowledge of the probability density functions $\rho$.
Such feature of RL algorithms leads to huge potential to optimize the agents in a model-free manner.

\subsection{Actor-Critic and Multi-agent Actor-Critic}
\label{sub:maacf}
In order to accomplish the reinforcement learning task, a group of RL algorithms called actor-critic algorithms are becoming popular in the recent years for their high sample efficiency and stability, such as PPO\cite{schulman2017proximal}, A3C\cite{mnih2016asynchronous}, DDPG\cite{silver2014deterministic}, and SAC\cite{haarnoja2018sacapplications}.
These algorithms utilize deep neural network to approximate an ``actor'', which generate actions with observations using policy $\pi$, and an ``critic'' which evaluate the policy using $Q^\pi$ or $V^\pi$.
By training the actor and critic alternatively, these algorithms could explore the environment efficiently and get high quality policies. The constrained version of SAC is also developed in \cite{8944292}.

\begin{figure}[h]
  \centering
  \includegraphics[width=0.95\linewidth]{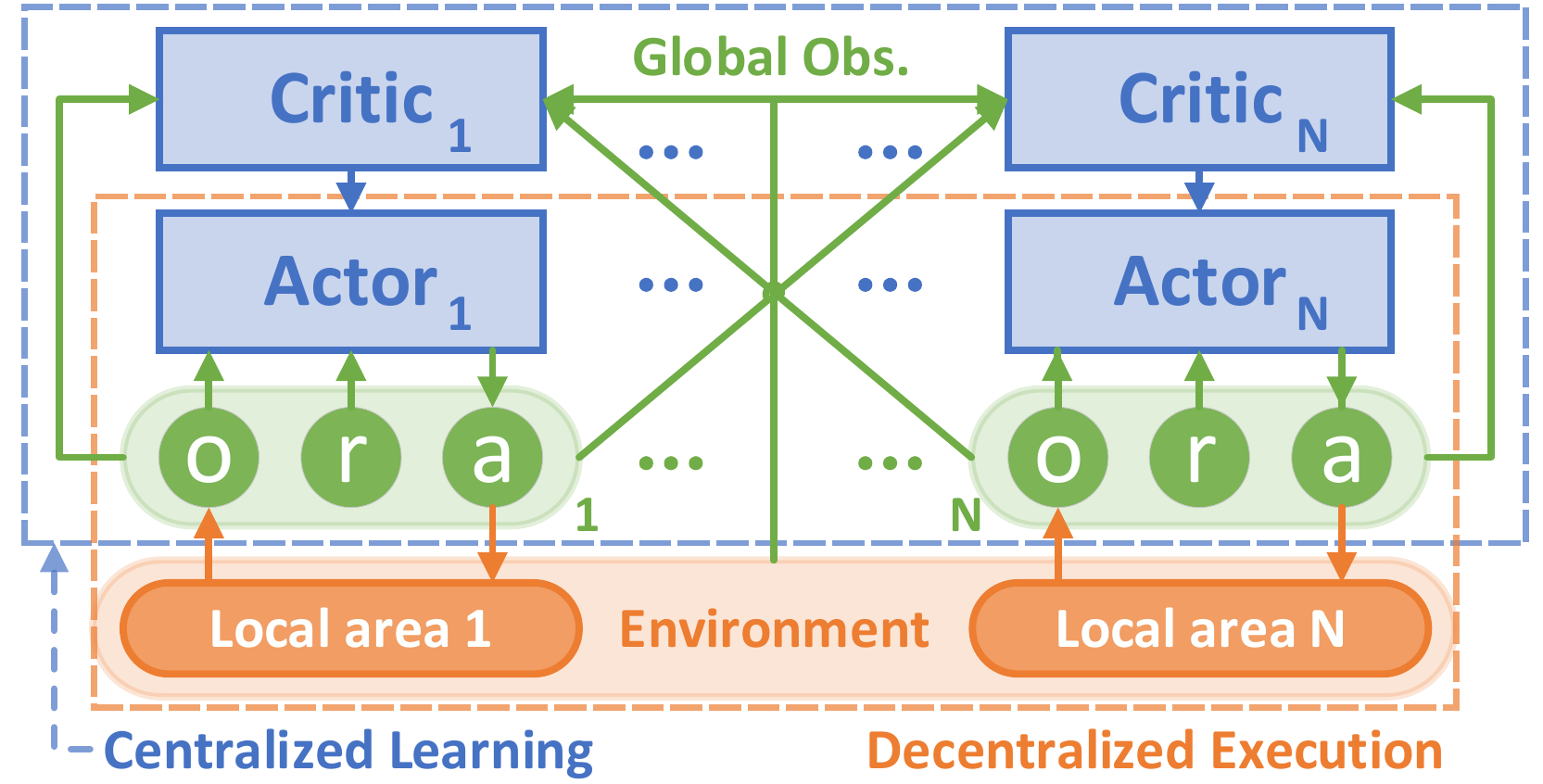}
  \caption{Multi-agent reinforcement learning system for VVC of ADNs.}
  \label{fig:marl}
\end{figure}

For such multi-agent environments, separately adopting traditional RL algorithms for each agent is poorly suited because the environment is non-stationary from the perspective of each individual agent.
In this paper, we follow the multi-agent actor-critic framework in \cite{foerster2016learning,NIPS2017_7217} to cope with the inherent non-stationary challenges of multi-agent environments. The architecture of the multi-agent RL system is illustrated in \cref{fig:marl}.
Both a critic and a local actor are constructed for each agent.
At training time, the critics are allowed to use global information, including all observations and actions, to build its own evaluation of the global environment characteristics.
The local actors are trained with the corresponding critic with the knowledge of other actors since we consider a cooperative setting in this paper.
After training is complete, the local actors are deployed and make decisions in a decentralized manner using only the local information.

However, previous work is not intended for online controlling and acts in an offline training and online application mode.
In our DRL-based VVC algorithm, the most important task is to utilized online learning and control to adaptively operates ADNs.
So in \cref{sub:framework}, we propose an online multi-agent learning and decentralized control framework (OLDC) with totally asynchronous sampling, training and application, which fully preserves the advantage of OLDC in the online stage.

\section{Methods}
\label{sec:methods}
\begin{figure}[htbp]
  \centering
  \includegraphics[width=\linewidth]{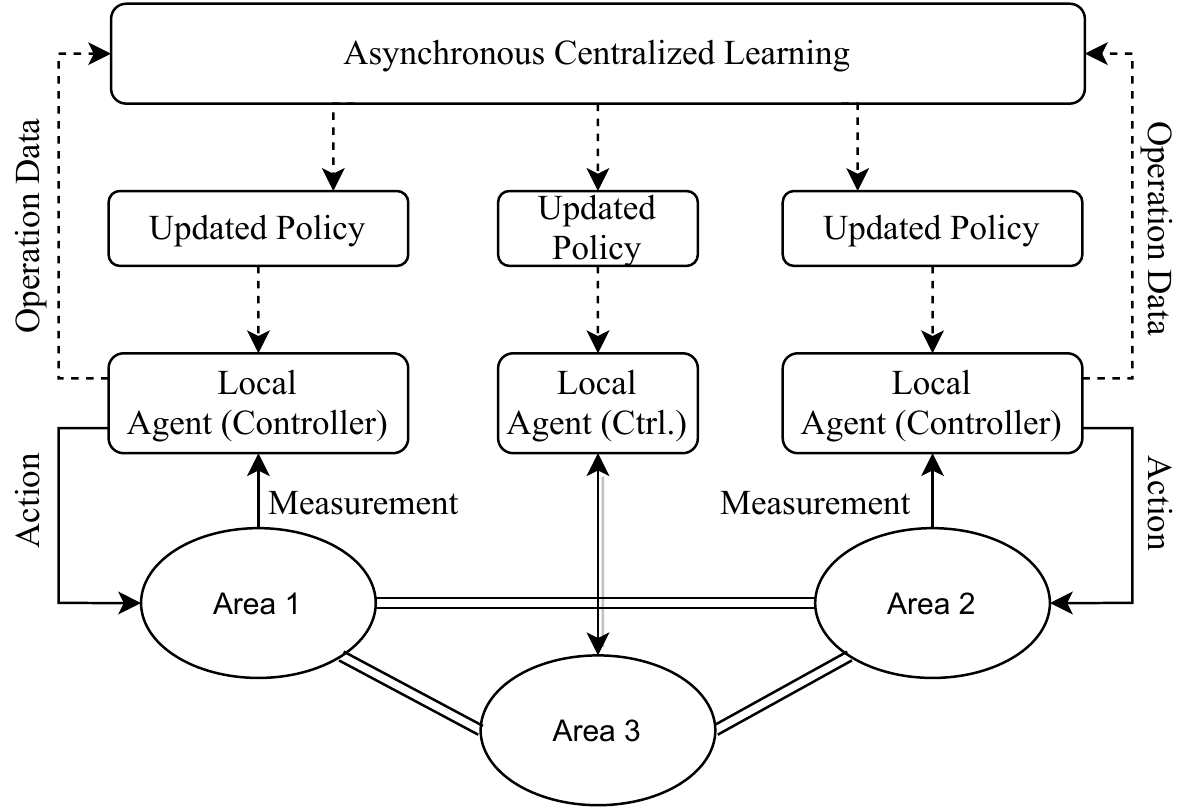}
  \caption{Structure diagram of the proposed OLDC. In the control process, each agent does not communicate at all; but asynchronously, the data is sent and computed in the control center, and the policies are updated, which does not affect the control process.}
  \label{fig:oldc}
\end{figure}

In this section, we innovate an online multi-agent reinforcement learning method to solve the VVC problem formulated as a MG.
Since the method is carried out online, the safety, efficiency and optimality are the critical concerns to address in the real world problem.
Firstly, the VVC problem is formulated into CMG.
Then, we develop an innovated off-policy multi-agent algorithm called MACSAC in \cref{sub:macsac}, which improves the safety and efficiency of the existing algorithms.
Finally, based on the off-policy nature of MACSAC, we propose OLDC as an online multi-agent actor-critic framework with totally asynchronous sampling, learning and application in \cref{sub:framework}, which is also capable with other off-policy algorithms. The structure diagram is shown in \cref{fig:oldc}, which emphasizes the decentralized nature of the control process and the asynchronous nature of the centralized learning part.

\subsection{VVC Formulation in Constrained Markov Game}
The VVC problem of ADNs is formulated as CMG with their natural features.
The detailed VVC problem settings are given in the supplemental file \cite{zzsupple} due to page limitation.
The specific definitions of state space, action space and reward function are designed as follows.

\subsubsection{State Space}
The state of CMG $s\in\mathcal{S}$ is defined as a vector $s=(\mathbf{P},\mathbf{Q},\mathbf{V},t)$.
Here $\mathbf{P},\mathbf{Q}$ is the vector of nodal active/reactive power injections $P_j, Q_j(\forall j \in \mathcal{N})$,
$\mathbf{V}$ is the vector of voltage magnitudes $V_j(\forall j \in\mathcal{N})$.
$t$ is the time step in each episode.

\subsubsection{Observation Spaces}
The local observations of each agent are selected according to the local measurements.
In this paper, $o_i\in\mathcal{O}_i$ is defined as $(\mathbf{P}_i,\mathbf{Q}_i,\mathbf{V}_i,\mathbf{P}^e_i,\mathbf{Q}^e_i)$,
where $\mathbf{P}_i,\mathbf{Q}_i$ is the vector of $i$th area's nodal active/reactive power injections $P_j, Q_j(\forall j \in \mathcal{N}_i)$;
$\mathbf{V}$ is the vector of $i$th area's voltage magnitudes $V_j(\forall j \in\mathcal{N}_i)$;
$\mathbf{P}^e_i,\mathbf{Q}^e_i$ is the vector of outlet powers of $i$th area.

\subsubsection{Action Spaces}
For each agent $i$, the action space $\mathcal{A}_i$ is constructed with all the controllable reactive power resources in $i$th area, including PV inverters and SVCs.
That is, $\mathcal{A}_i = \{Q_{Gj}, Q_{Ck}\}, j \in \mathcal{N}_\text{IBi}, k\in \mathcal{N}_\text{CDi}$, which is similar to \cite{8960272}.

\subsubsection{Reward and Cost Functions}
In the classic RL algorithms, the reward is designed to be a function of previous observations.
In this paper, the rewards of agents are calculated in the coordinator, so all observations are available to the reward functions.
Since the objectives are to minimize active power loss and mitigate voltage violations, the reward functions and cost functions are defined as \cref{eq:r} and \cref{eq:rc}.
$\beta_i$ is the cooperative index of agent $i$, which describes the willingness of the agent to optimize the welfare for global system rather than itself.
\begin{gather}
  \label{eq:r}
  r_{i,t} = R_P(t) =  \sum_{i\in[1,N]}[ P^e(\mathcal{N}_i)-\sum_{j\in\mathcal{N}_i}P_{j}(t) ]\\
  \label{eq:rc}
  r^c_{i,t} = R_V(\mathcal{N}_i, t) + \beta_i R_V(\mathcal{N}, t)
\end{gather}

The index functions $R_P$ and $R_V$ can be evaluated in the coordinator for any collection of nodes $\mathcal{N}_i$ at time step $t$.
\begin{equation}
    R_V(\mathcal{N}_i,t) = \sum_{j\in\mathcal{N}_i}\left[[V_j(t)-\overline{V}]_{+}^2 + [\underline{V}-V_j(t)]_{+}^2\right]
  \end{equation}

Here, $[\cdot]_+$ is the rectified linear unit function defined as $[x]_+=\max(0, x)$.
We have $R_V(t) \geq 0$ where the equality holds if and only if all voltage magnitudes satisfy the voltage constraints.
Note $R_V$ as voltage violation rate (VVR) since it is assigned according to the 2-norm of voltage magnitude violations.
We use VVR instead of the amount of violated nodes like \cite{8944292} because the voltage violations are usually severe in the ADNs and the regulation capacity may be not enough to eliminate all violations in some scenarios.
In such scenarios, VVR serves as a much smoother index and can effectively mitigate the voltage violations.

\subsection{Multi-agent Constrained Soft Actor-Critic}
\label{sub:macsac}
To improve the safety and efficiency of the existing multi-agent RL algorithms, we propose MACSAC in this subsection. As space is limited, the detailed derivation of MACSAC and practical skills are provided in the supplemental file \cite{zzsupple}.

First of all, with the formulation of CMG for VVC in \cref{sec:prelim}, the multi-agent RL problem is reformulated as \cref{eq:problem,eq:cons_a,eq:cons_h,eq:cons_r} for each agent $i$ locally. Here, \cref{eq:problem} is the original RL objective; \cref{eq:cons_a} is the action constraint, where $\underline{a_i}$ and $\overline{a_i}$ is the lower and upper bound of $a_i$; \cref{eq:cons_h} is the entropy constraint from \cite{haarnoja2018sacapplications}, where $\mathcal{H}_i$ is the lower bound of $\pi_i$'s entropy; \cref{eq:cons_r} is the state constraint of our CMG, i.e., the expected discount sum of VVR.
\begin{gather}
  \label{eq:problem}
  \max_{\pi_i} J^0_i(\pi_i) = \mathop{\mathbb{E}}\limits_{\tau\in\rho_\phi}\left[\sum_{t=0}^T \gamma^t r_{i,t}\right], \quad s.t.\\
  \label{eq:cons_a}
  \underline{a_i} \leq a_i \leq \overline{a_i},\\
  \label{eq:cons_h}
  \mathop{\mathbb{E}}\limits_{(o_{i,t}, a_{i,t}) \in \rho_\pi}\left[ -\log\left( \pi_i(a_{i,t}\left|o_{i,t}\right.) \right) \right] \geq \mathcal{H}_i,\,\forall t, \\
  \label{eq:cons_r}
  \overline{J}^c_i\geq J^c_i(\pi_i)=\mathop{\mathbb{E}}\limits_{\tau\in\rho_\phi}\left[\sum_{t=0}^T \gamma^t r_{i,t}^c\right].
\end{gather}

For the action constraint \cref{eq:cons_a}, it has already been included in the action spaces' definition.
As usual, we adapt Lagrange relaxation here to handle constraints \cref{eq:cons_h,eq:cons_r}.
Multipliers $\alpha_i$ and $\lambda_i$ are introduced for \cref{eq:cons_h} and \cref{eq:cons_r} respectively.
Note that $(\alpha_i,\mathcal{H}_i)$ and $(\lambda_i, \overline{J}^c_i)$ are two pairs of variables.
In each pair, if one variable is considered as a hyperparameter, the other one can be determined via iterations.
Since the physical meaning of $\overline{J}^c_i$ is clear, we select $\alpha_i$ and $\overline{J}^c_i$ as hyperparameters.
Hence, the problem is refined as $\max\limits_{\pi_i} \min\limits_{\lambda_i} J_i + \lambda_i \left[ \overline{J}^c_i - J^c_i(\pi_i) \right]$, where $J_i = \mathop{\mathbb{E}}\limits_{\tau\in\rho_\phi}\left[\sum_{t=0}^T \gamma^t r_{i,t} - \alpha_i \log\left( \pi_i(a_{i,t}\left|o_{i,t}\right.) \right) \right]$.
In the traditional RL-based algorithms, the voltage constraints are penalized directly in the reward, which means the multipliers $\lambda_i$ here are designed to be a given penalty hyperparameter. With an inappropriate penalty hyperparameter, the voltage constraints can not be satisfied or lead to unpreferred convergence. In this paper, the dynamic update of multiplier $\lambda_i$ guarantees the safety of the proposed algorithm.

\subsubsection{Preparation}
The actors optimize the policies $\pi_{\theta_i}, i\in[1, N]$ with parameters $\theta_i, i\in[1, N]$ according to the optimization problem above. In MADDPG, $\pi_i$ is defined as a deterministic map from $\mathcal{O}_i$ to $\mathcal{A}_i$, but faces overfitting problem and shows undesirable instability. Inspired by \cite{haarnoja2018sacapplications}, $\pi$ is defined as a probability distribution $\pi_i(\cdot|o_{i,t})$ here in a stochastic manner. Since directly optimization of a distribution is hard to implement, the policies $\pi_i$ is reparameterized as
\begin{equation}
\tilde{a}_{\theta_i}(o_i, \xi_i) = \tanh\left( \mu_{\theta_i}(o_i) + \sigma_{\theta_i}(o_i) \odot \xi_i \right), \xi_i \sim \mathcal{N}(0, I)
\end{equation}
where $\mu_{\theta_i}, \sigma_{\theta_i}$ is the mean and standard deviations approximated by neural networks.

In order to quantify the policies, the state-action value functions $Q_i^\pi(\mathbf{x},\mathbf{a})$ are defined in \cref{eq:qi} for $J_i$.
$Q_i^\pi(\mathbf{x},\mathbf{a})$ is representing the expected discounted reward after taking action $\mathbf{a}$ under observation $\mathbf{x}$ with the policy $\pi$.
Here, $\tau \sim \pi$ is the trajectory when applying $\pi$; $\pi$ is noted for all $\pi_{\theta_i},i\in[1,N]$; $\mathbf{x}$ is all observations $(o_0, \dots, o_N)$; $\mathbf{a}$ is all actions $(a_1, \dots, a_N)$. At every time step $t$, we store $\{\mathbf{x},a,r,\mathbf{x}'\}_t$ in the experience replay buffer $D$, and then learn the critics and actors alternatively as follows.
\begin{equation}
  \label{eq:qi}
  \begin{split}
    &Q_i^\pi(\mathbf{x},\mathbf{a}) \doteq \\
    &\mathop{\mathbb{E}}\limits_{\tau \sim \pi}\left[\sum_{t=0}^T \gamma^t r_{i,t} - \alpha_i\sum_{t=1}^T \gamma^t \log\pi_i(\cdot|\mathbf{x}_t) \left| \mathbf{x}_0 = \mathbf{x}, \mathbf{a}_0 = \mathbf{a}\right. \right]
  \end{split}
\end{equation}
From the definition, the only difference between each $Q_i^\pi$ is $r_{i,t}$ and $\pi_i$.
In the rest of MACSAC, we use neural networks $Q_{\phi_i}^\pi$ to approximate the actual $Q_i^\pi$.

As for the state constraint term $J^c_i$, similar state-action value functions $Q^{c, \pi}_i$ are defined by substituting $r_{i,t}$ with $r^c_{i,t}$ in \cref{eq:qi}.

\subsubsection{Learning the critics}
As defined in \cref{eq:qi}, we learn centralized critics with all observations and actions instead of learn local ones separately. Such manner can cope with the non-stationary problem from the perspective of any individual agents.
Since in this paper the agents are cooperative, the policies of others are available when training a certain critic.

Using Bellman equation, we could approximate the current state-action value with the expectation of all possible next state and corresponding actions with $\pi$. That is, 
\begin{equation}
  \label{eq:bellman}
  \begin{split}
    &Q_{\phi_i}(\mathbf{x}, a_1, \dots, a_N) \approx \mathop{\mathbb{E}}\limits_{\mathbf{x},a,r,\mathbf{x}'}\left[ y_i \right] \\
    &y_i = r_i + \gamma \left[ 
      Q_{\hat\phi_i}(\mathbf{x}',\tilde{a}'_1,\dots,\tilde{a}'_N) 
      - \alpha_i \log \pi_{\theta_i}
      (\tilde{a}'_i\left|o_i'\right.) \right]
  \end{split}
\end{equation}
where $\tilde{a}'_i \doteq \tilde{a}_{\theta_i}(o_i', \xi_i)$; $\hat\phi_i$ is the delayed parameters for $\phi_i$ and is updated using $\hat\phi_i \leftarrow \eta \phi_i + (1 - \eta)\hat\phi_i$.

Hence, the training of $\phi_i$ is to minimize the loss $\mathcal{L}(\phi_i) = \mathop{\mathbb{E}}\limits_{\mathbf{x},a,r,\mathbf{x}'}\left[ \left( Q_{\phi_i}(\mathbf{x}, a_1, \dots, a_N) - y_i \right)^2 \right]$.

Similarly, we calculate the approximated value for $Q^c_{\varphi_i}$ as $y^c_i = r^c_i + \gamma Q^c_{\hat\varphi_i}(\mathbf{x}',\tilde{a}'_1,\dots,\tilde{a}'_N)$, and update $\varphi_i$ by minimizing the loss $\mathcal{L}(\varphi_i) = \mathop{\mathbb{E}}\limits_{\mathbf{x},a,r,\mathbf{x}'}\left[ \left(Q^c_{\varphi_i}(\mathbf{x}, a_1, \dots, a_N) - y^c_i \right)^2 \right]$.

\subsubsection{Learning the actors}
With the definition of critics, the optimization problem of actors is transformed from maximizing $J_i$, which is hard to get, to \cref{eq:piproblem} with approximated $Q_{\phi_i}$ and $Q^c_{\varphi_i}$.
\begin{equation}
  \label{eq:piproblem}
  \begin{split}
    \max_{\theta_i} &\mathop{\mathbb{E}}\limits_{\mathbf{x}\sim D}\left[ Q_{\phi_i}(\mathbf{x}, \tilde{a}_1,\dots,\tilde{a}_N)
    - \alpha_i \log \pi_{\theta_i} (\tilde{a}_i \left| o_i \right.) \right] \\
    s.t.\quad &\mathop{\mathbb{E}}\limits_{\mathbf{x}\sim D}\left[ Q^c_{\varphi_i}(\mathbf{x}, \tilde{a}_1,\dots,\tilde{a}_N)\right] \leq \overline{J}^c_i
  \end{split}
\end{equation}
where $\tilde{a}_i \doteq \tilde{a}_{\theta_i}(o_i, \xi_i)$.

The Lagrange function $L(\theta_i,\lambda_i)$ is derived for \cref{eq:piproblem} as,
\begin{equation}
  \label{eq:lagrange}
  \begin{split}
    L(\theta_i,\lambda_i) = &\mathop{\mathbb{E}}\limits_{\mathbf{x}\sim D}\left[ Q_{\phi_i}(\mathbf{x}, \tilde{a}_1,\dots,\tilde{a}_N)
    - \alpha_i \log \pi_{\theta_i} (\tilde{a}_i \left| o_i \right.) \right] \\
    &+ \lambda \left[ \overline{J}^c_i - 
    \mathop{\mathbb{E}}\limits_{\mathbf{x}\sim D}\left[ Q^c_{\varphi_i}(\mathbf{x}, \tilde{a}_1,\dots,\tilde{a}_N)\right] \right]
  \end{split}
\end{equation}

Hence, the dual problem for $\theta_i$ and $\lambda_i$ is $\max\limits_{\theta_i} \min\limits_{\lambda_i} L(\theta_i, \lambda_i)$. In MACSAC, we update $\theta_i$ as $\theta_i + \sigma^\theta_i\nabla_{\theta_i} L(\theta_i,\lambda_i)$, and update $\lambda_i$ as $\left[\lambda_i - \sigma^\lambda_i \nabla_{\lambda_i} L(\theta_i,\lambda_i) \right]_{+}$.

\begin{algorithm}
  \caption{Multi-agent Constrained SAC}
  \SetKwProg{Parallel}{foreach}{ do in parallel}{end}
  \label{alg:macsac}
  Initialize experience pool $D$, policy and value function approximators' parameter vectors\;
  \ForEach{
    episode
  }{
    \ForEach{
      environment step $t$
    }{
      \Parallel{
        agent $i$
      }{
        Locally observe $o_{i,t}$\;
        $a_{i,t}=\tilde{a}_{\theta_i}(o_i, \xi_i), \,\, \xi_i \sim \mathcal{N}(0, I)$\;
        Feed $a_{i,t}$ to the environment and get reward $r_{i,t}$ and next observation $o_{i,t+1}$\;
      }
      $\mathcal{D}\leftarrow\mathcal{D}\cup\{(\mathbf{x}_t,\mathbf{a}_t,\mathbf{r}_t,\mathbf{x}_{t+1})\}$\;
      \Parallel{
        agent $i$
      }{
        Sample a batch $B_{i,t}$ from $D$\;
        Update $Q_{\phi_i}$: $\phi_i\leftarrow \sigma_i\nabla_{\phi_i}\mathcal{L}(\phi_i)$\;
        Update $Q^c_{\varphi_i}$: $\varphi_i\leftarrow \sigma_i\nabla_{\varphi_i}\mathcal{L}(\varphi_i)$\;
        Update $\pi_{\theta_i}$: $\theta_i \leftarrow \theta_i + \sigma_i\nabla_{\theta_i} L(\theta_i,\lambda_i)$\;
        Update $\lambda_i$: $\lambda_i \leftarrow \left[ \lambda_i - \sigma_i \nabla_{\lambda_i} L(\theta_i,\lambda_i) \right]_+$\;
        $\hat\phi_i \leftarrow \eta \hat\phi_i + (1 - \eta)\phi_i$\;
        $\hat\varphi_i \leftarrow \eta \hat\varphi_i + (1 - \eta)\varphi_i$\;
      }
    }
  }
\end{algorithm}

The algorithm of MACSAC is shown in \cref{alg:macsac}.
Compared to the state-of-art multi-agent RL algorithm MADDPG \cite{NIPS2017_7217},
our MACSAC a) utilizes stochastic policies instead of deterministic policies for each agent and follow the maximum-entropy training in \cite{haarnoja2018sacapplications},
which avoids overfitting to local optimal polices and gains significantly higher sample efficiency and stability,
and b) introduces constraints for each agent and solve CMG instead of MG,
which guarantees voltage safety explicitly.
Also, both MADDPG and MACSAC are off-policy actor-critic algorithms, since we do not have any assumption with the order of samples or samples' original policy.
It means that the sampling policies, which are executed locally, are not required to be the latest policies.
Such feature inspires us to come up with OLDC as follows. 

\subsection{Online Centralized Training and Decentralized Execution Framework}
\label{sub:framework}
\begin{figure}[htbp]
  \centering
  \includegraphics[width=\linewidth]{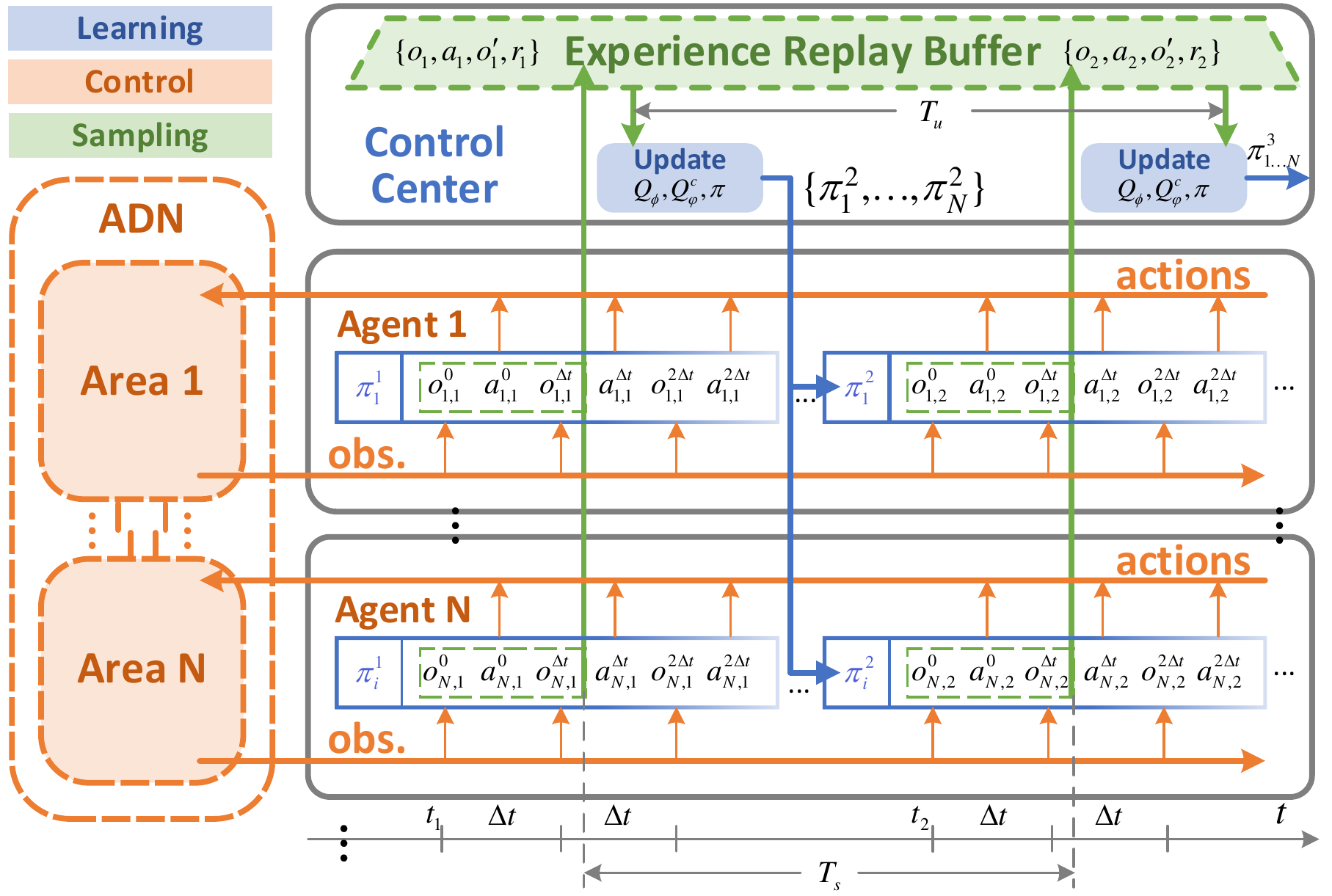}
  \caption{The proposed OLDC framework with totally asynchronous sampling (green), learning (blue) and application (orange).}
  \label{fig:main}
\end{figure}

With the physical structure shown in \cref{fig:proposed}, we propose OLDC to carry out MACSAC online with high efficiency. The detailed diagram of OLDC is illustrated in \cref{fig:main}.
Note that in OLDC, sampling (green), learning (blue) and application (orange) are totally asynchronous.

\subsubsection{Timing}
In the bottom of \cref{fig:main}, a timeline is built for all agents and the centralized server.

As the orange part, in every time gap $\Delta t$, each agent $i$ a) get the local measurement $o_i$, b) generate the action $a_i$ with local policy $\pi_i$ as $a_i\sim \pi_i(\cdot|o_i)$, and c) send $a_i$ to local controlled devices. Note that the lower bound of $\Delta t$ depends on the measurements, computation of $\pi_i$, and devices. In this control process, no centralized communication is needed and all computations are carried out locally in a decentralized manner. Since we consider high-speed measurements and devices, and $\pi_i$ is reparameterized as $\tilde{a}_{\theta_i}(o_i, \xi_i)$ with neural networks and can be fast evaluated, $\Delta t$ can be relatively small.

Asynchronously, the samples got in every $T_s$ is uploaded to the experience replay buffer on the server as the green part. Because of relatively slow communication, $T_s$ is much greater than $\Delta t$. However, the sampling process would not delay the actual control speed, since all application is carried out locally as above.

Also asynchronously as the blue part, the training of agents is carried out every $T_u$: batch of samples $B\in D$ is randomly selected to train the critics and actors using \cref{eq:bellman,eq:lagrange}, and the updated policies $\pi$ are sent to the agents.
Since the communication is relatively slow and computations is relatively heavy, $T_u$ is also much greater than $\Delta t$.
Note that the training process would not delay the application or sampling; also, the samples are selected from the experience replay buffer, so the training is not directly affected by $T_s$.

\subsubsection{Communication and Computation}
OLDC is robust to communication and computation conditions.
In the application process, local controller only evaluates a small neural network from local measurements $o_i$ to $a_i$ for local devices with little computation burdens and no communications are needed with other controllers or upper control center.
Most computations of MACSAC are carried out on the centralized server with abundant resources.

OLDC could choose to upload any proper numbers of samples in every $T_s$ considering the communication conditions.
Without loss of generality, one sample is drawn in \cref{fig:main} with dashed green box.
Also, even if the communication to the server is unstable and some samples were lost, they could be ignored safely.

\subsubsection{Exploration and Exploitation}
For data-driven algorithms like MACSAC, the balance of exploration and exploitation is extraordinary important. In MACSAC, bigger multiplier $\alpha_i$ will results in higher entropy level, which means $\pi_i$ is more stochastic and explore the environment better.
However, the exploration will sacrifice the exploitation, i.e., optimality and performance. In the practical application, a smaller $\alpha_i$ is preferred as long as the convergence and learning efficiency are satisfactory.

Hence, OLDC provides another way to balance exploration and exploitation.
Suppose we upload $m$ samples in every $T_s$, which means $0 \leq m < T_s/\Delta t$.
Since other samples are not uploaded or used in training, we can carry out the policy in a deterministic manner, that is, $\tilde{a}_{\theta_i}(o_i, 0) = \tanh\left( \mu_{\theta_i}(o_i)\right)$ instead of $\tilde{a}_{\theta_i}(o_i, \xi_i), \xi_i \sim \mathcal{N}(0, I)$.
To be brief, only the actions of samples which are meant to upload should explore stochastically in OLDC.
With smaller $m$, the exploration is weaker and exploitation is stronger.
Moreover, $m$ and $T_s$ can be changed online to manually control the learning process or even stop learning with $m=0$. With a proper tuned $m$ and $T_s$, the efficiency of MACSAC can be dramatically improved in the online application.

\subsubsection{Special Case}
As a special case, OLDC is also capable with single-agent actor-critic RL, i.e., $N=1$.
The sampling, training and execution are still asynchronous if needed.

With extraordinary efficiency and robustness to various computing and communication conditions, OLDC is a practical and suitable framework for online (MA)RL application in the power system, especially for multi-agent RL-based VVC in the ADNs.

\section{Numerical Study}
\label{sec:numerical}
In this section, numerical experiments are conducted to validate the advantage of the proposed OLDC and MACSAC over some popular benchmark algorithms including DRL algorithms and optimization-based algorithms.
Multi-agent RL environments are built of steady-state power systems under the scheme of the toolkit Gym \cite{brockman2016openai}.
Both the balanced 33-bus test feeder \cite{25627} and 141-bus test feeder \cite{KHODR20081192} are adapted as ADNs. The balanced power flow equations are solved to simulate the ADNs.
All of the algorithms are implemented in Python, while the DRL-based algorithms utilize deep learning framework PyTorch, and the optimization-based methods utilize Casadi \cite{Andersson2019} and Ipopt.Experiments are run on a MacBook Pro with 16GB memory and 3.1GHz dual-core Intel i5 CPU.
In the 33-bus case, there are three PV inverters and one SVC, which are assumed as four stations.
In the 141-bus case, we have 13 PV inverters, 5 SVCs and 5 stations.
The base of powers is set as 1 MVA.
Detailed simulation configuration and load/generation profiles are given in the supplemental file \cite{zzsupple}.
\subsection{Proposed and Baseline Algorithms Setup}
In the following experiments, the proposed MACSAC is implemented with our OLDC.
For the benchmark algorithm, we adapt the state-of-art MADDPG \cite{NIPS2017_7217} as a multi-agent RL baseline, and CSAC from \cite{8944292} as a centralized RL baseline.
An optimization-based algorithm with SOCP relaxation is implemented with oracle models (VVO), which could serve as a benchmark of theoretically best performance.
VVO with approximated models and practical considerations is treated as the model-based benchmark called approximated VVO (AVVO).
The algorithm hyper-parameters for RL algorithms are listed in the supplemental file \cite{zzsupple}.

Due to the stochastic property of DRL-based algorithms, we use 3 independent random seeds for each group of experiment, whose mean values and error bounds are presented in the figures as solid lines and filled areas.
\subsection{Algorithm Convergence and Efficiency with Ideal Simulation}
To verify the convergence and efficiency of the proposed MACSAC, we first conduct an ideal centralized experiment with the RL algorithms, in which all RL algorithms do not consider the speed of communication, that is, CSAC in a centralized manner and MACSAC / MADDPG in OLDC ($T_u = T_s = 1$) can execute the policies in every time step.
During the execution, all samples are uploaded to the experience replay buffer.
In this first experiment, all stochastic explorations are carried out in an identical copy of our simulated system, thus policies are free of noisy explorations in the our testing algorithms for now. Note that though such noise-free scenario is actually not realistic in practice, the results of which are informative for making it more explicit to  compare the convergence and efficiency of RL and multi-agent RL approaches.

\begin{figure}[htbp]
  \centering
  \includegraphics[width=0.95\linewidth]{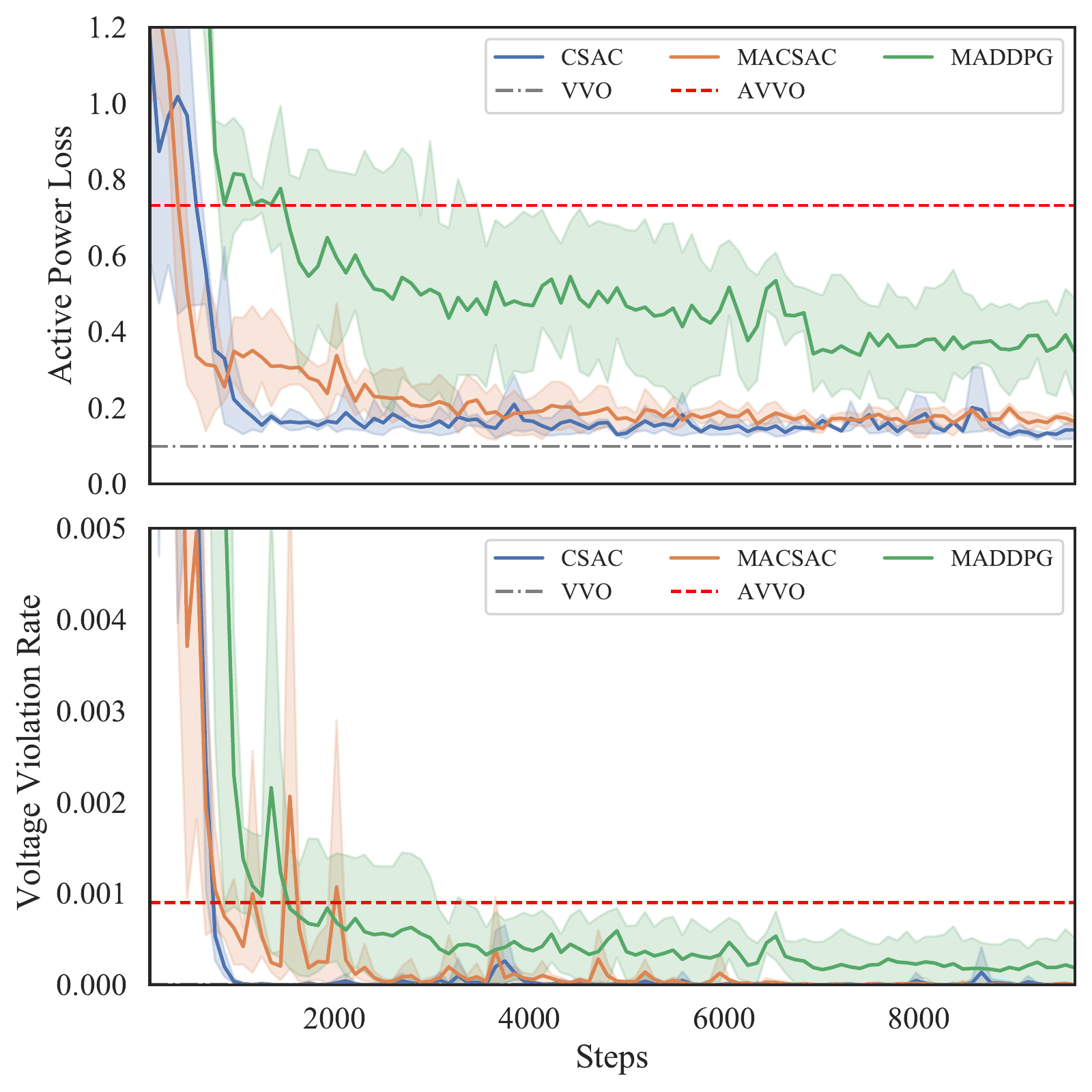}
  \caption{33-bus case results of MACSAC and benchmarks under the ideal scenario without communication delay and stochastic exploration in test.}
  \label{fig:all}
\end{figure}

\begin{figure}[htbp]
  \centering
  \includegraphics[width=0.95\linewidth]{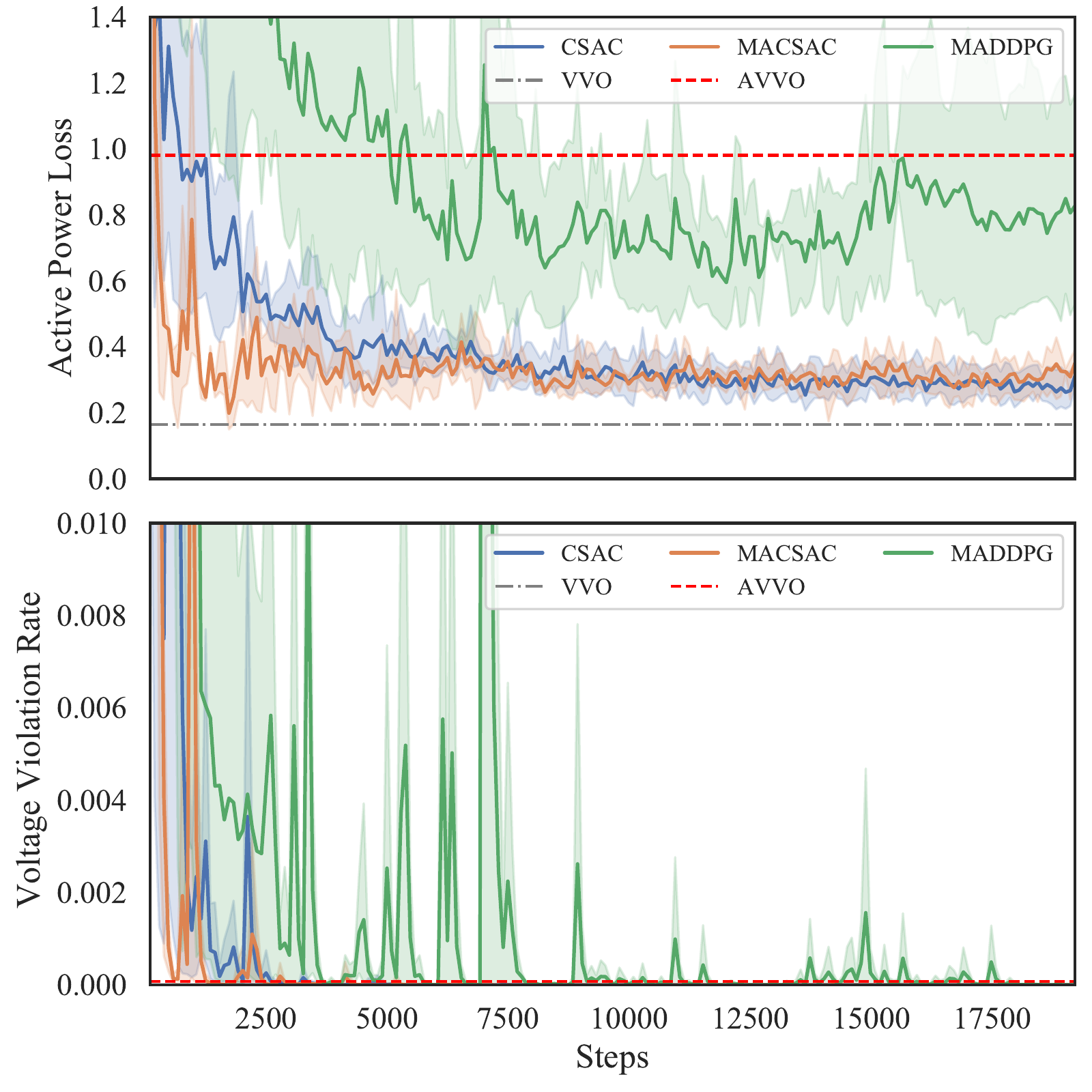}
  \caption{141-bus case results of MACSAC and benchmarks under the ideal scenario without communication delay and stochastic exploration in test.}
  \label{fig:all141}
\end{figure}

The step value of active power loss and VVR during the training process are shown in \cref{fig:all,fig:all141}.
The model-based benchmark VVO is also tested with results averaged across load/generation profile since it is deterministic.

\begin{table}[h]
  \centering
  \begin{threeparttable}
    \caption{Quantified indices of the final episode in ideal scenario}
    \label{tbl:ideal}
    \begin{tabular}{@{}crrrrr@{}}
      \toprule
      \multicolumn{1}{l}{\multirow{2}{*}{Test system}} & \multicolumn{1}{r}{\multirow{2}{*}{Algorithm}} & \multicolumn{2}{c}{$P_\text{loss}/\text{MW}$}    & \multicolumn{2}{c}{VVR}      \\ \cmidrule(l){3-6}
      & \multicolumn{1}{c}{}    & \multicolumn{1}{c}{Mean} & Std. & Mean & Std. \\ \midrule
      \multirow{5}{*}{33-bus sys.}
                        & CSAC       & 1.43e-01 & 2.46e-02 & 5.38e-06 & 5.20e-06 \\
                        & MADDPG     & 3.47e-01 & 1.34e-01 & 1.88e-04 & 2.80e-04 \\
                        & MACSAC     & 1.64e-01 & 1.22e-02 & 4.19e-06 & 4.13e-06 \\
                        & AVVO\tnote{+} & 7.32e-01 & -        & 9.02e-04 & -        \\
                        & VVO\tnote{o} & 9.83e-02 & -   & 8.51e-06 & -  \\ \midrule
      \multirow{5}{*}{141-bus sys.}
                        & CSAC       & 3.01e-01 & 3.88e-02 & 0 & 0 \\
                        & MADDPG     & 8.27e-01 & 4.00e-01 & 2.02e-06 & 2.53e-06 \\
                        & MACSAC     & 3.51e-01 & 3.57e-02 & 0 & 0 \\
                        & AVVO\tnote{+} & 9.81e-01 & -        & 7.83e-05 & -        \\
                        & VVO\tnote{o} & 1.64e-01 & -   & 0 & -  \\ \bottomrule
    \end{tabular}
    \begin{tablenotes}
      \item[+] optimization-based VVO with the approximate model.
      \item[o] ideal optimal solution using perfect ADN model.
    \end{tablenotes}
  \end{threeparttable}
\end{table}

The first important observation from \cref{fig:all,fig:all141} is that both CSAC and MACSAC converge to a lower active power loss than the optimization-based method AVVO without oracle parameters, which reveals the advantage of DRL-based algorithms over such parameter-sensitive optimization method regarding VVC problem.
On the other hand, though the oracle VVO attains the minimum of active power loss theoretically once given all true parameters, DRL-based algorithms could closely approach it after certain iterations, as depicted in the figure.

Only using local measurements during application for each agent, MACSAC has achieved similar performance as the centralized algorithm CSAC, which in comparison utilizes global measurements during application, even in an ideal centralized scenario advantageous for the latter. Such results strongly support the fact that the CMG formulation and OLDC-like learning framework is valid for VVC in ADNs.

Also, MACSAC outperforms MADDPG obviously regrading active power loss and VVR in limited steps as \cref{fig:all,fig:all141} shows.
In fact, this significant improvement in MACSAC compared to MADDPG is credited to the usage of maximum-entropy regularized stochastic policies rather than deterministic policies, since the latter could easily overfit the value functions and lead to extreme brittleness \cite{haarnoja2018sacapplications}.
Such features make MACSAC preferable in practice for multi-agent VVC, not only in this study but also in more complex potential tasks.
\subsection{Online Application Performance with Real-world Simulation}
\label{sub:online}
To simulate the online stage, practical considerations include:
a) communication speed is limited comparing to the control speed, so the centralized algorithm CSAC can generate actions every 8 steps;
b) exploration has to be performed on the real system;
and c) training and sampling can be performed every 8 steps.
With the stochastic explorations, all RL-based methods including CSAC, MADDPG and MACSAC have to keep a stochastic range around the policy outputs and put the stochastic action directly to the real system.
The performance would be affected comparing with the ideal scenarios above.
Since the original OLDC framework is not suitable for online learning, we implement both MACSAC and MADDPG under OLDC with $T_u = T_s = 8$ and $m=1$.
Note that VVO is still implemented in the ideal scenario to provide a lower bound reference.
\begin{figure}[htbp]
  \centering
  \includegraphics[width=0.95\linewidth]{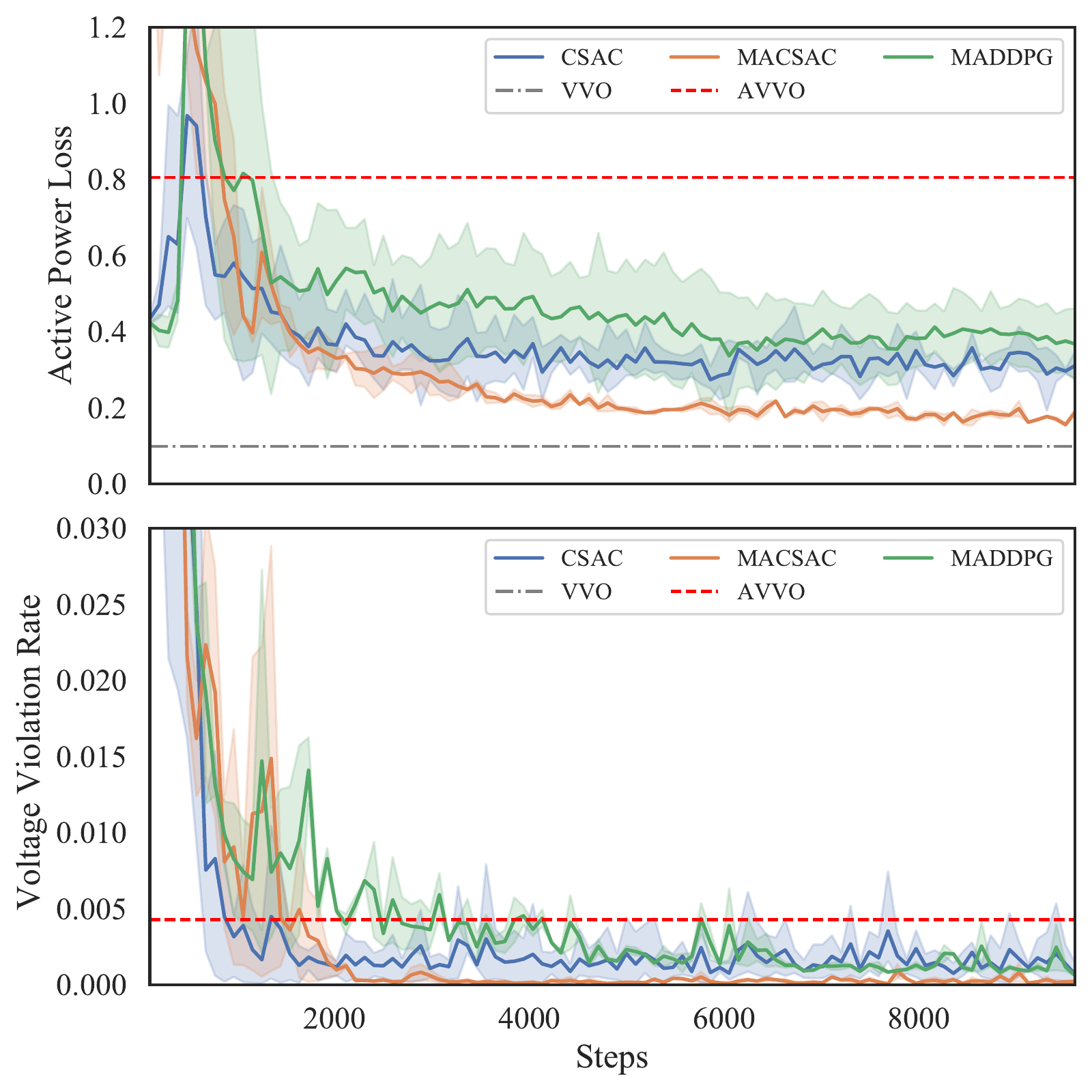}
  \caption{Online application performance with real-world simulation of 33-bus case.}
  \label{fig:delayed}
\end{figure}
\begin{figure}[htbp]
  \centering
  \includegraphics[width=0.95\linewidth]{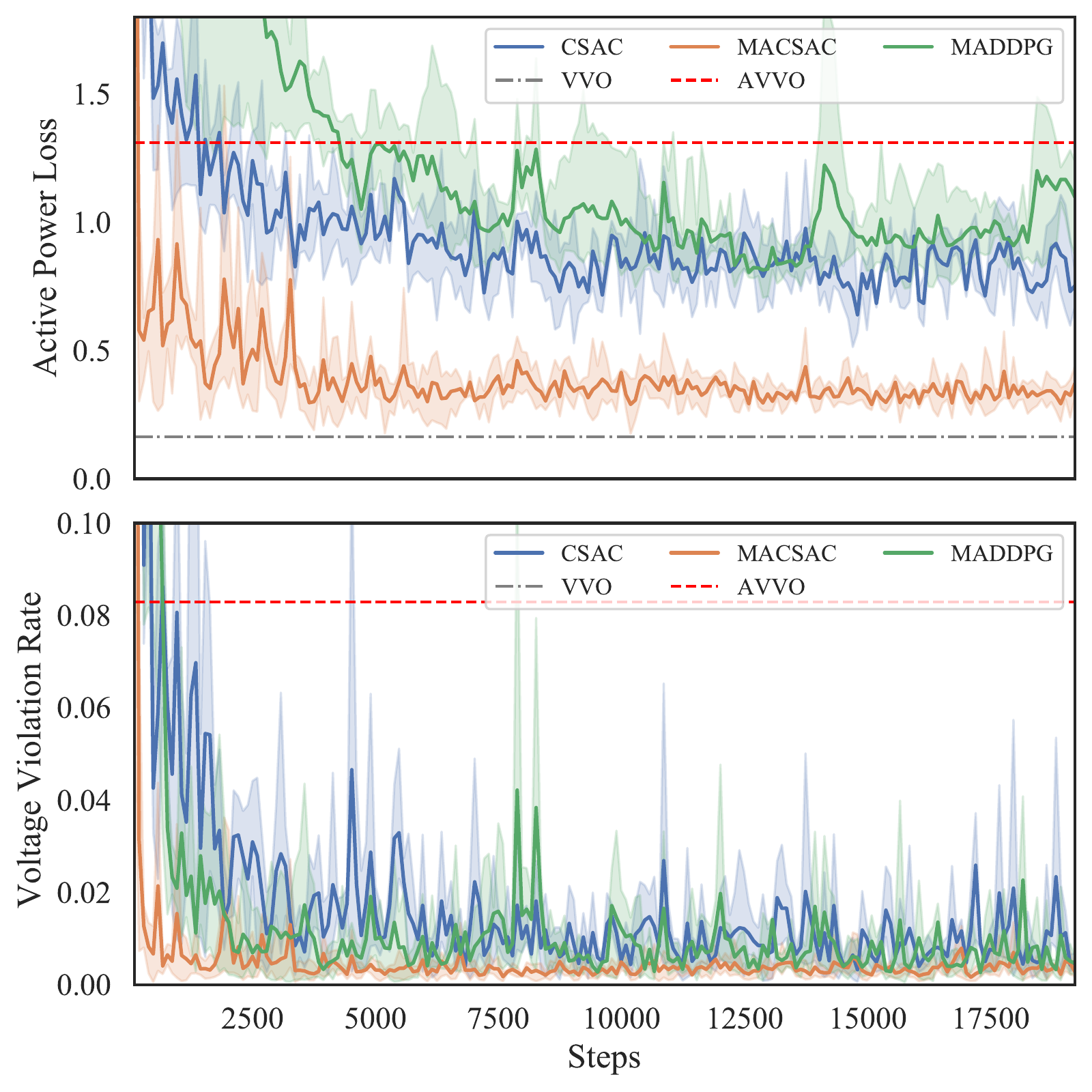}
  \caption{Online application performance with real-world simulation of 141-bus case}
  \label{fig:delayed141}
\end{figure}

\begin{table}[h]
  \centering
  \begin{threeparttable}
    \caption{Quantified indices of the final episode in online application}
    \label{tbl:online}
    \begin{tabular}{@{}crrrrr@{}}
      \toprule
      \multicolumn{1}{l}{\multirow{2}{*}{Test system}} & \multicolumn{1}{r}{\multirow{2}{*}{Algorithm}} & \multicolumn{2}{c}{$P_\text{loss}/\text{MW}$}    & \multicolumn{2}{c}{VVR}      \\ \cmidrule(l){3-6}
      & \multicolumn{1}{c}{}    & \multicolumn{1}{c}{Mean} & Std. & Mean & Std. \\ \midrule
      \multirow{5}{*}{33-bus sys.}
                        & CSAC       & 3.27e-01 & 2.81e-02 & 3.24e-04 & 6.73e-05 \\
                        & MADDPG     & 3.68e-01 & 1.32e-01 & 6.31e-04 & \textbf{5.31e-05} \\
                        & MACSAC     & \textbf{1.88e-01} & \textbf{1.08e-02} & \textbf{2.19e-04} & 1.55e-04 \\
                        & AVVO\tnote{+} & 8.06e-01 & -        & 4.29e-03 & -        \\
                        & VVO\tnote{o} & 9.83e-02 & -   & 8.51e-06 & -  \\ \midrule
      \multirow{5}{*}{141-bus sys.}
                        & CSAC       & 7.73e-01 & 9.02e-02 & 7.24e-03 & 3.31e-03 \\
                        & MADDPG     & 1.10e-00 & 1.98e-01 & 4.65e-03 & 1.98e-01 \\
                        & MACSAC     & \textbf{3.72e-01} & \textbf{6.24e-02} & \textbf{2.84e-03} & \textbf{9.67e-04} \\
                        & AVVO\tnote{+} & 1.30e-00 & -        & 8.29e-02 & -        \\
                        & VVO\tnote{o} & 1.64e-01 & -   & 0 & -  \\ \bottomrule
    \end{tabular}
    \begin{tablenotes}
      \item[+] optimization-based VVO with the approximate model.
      \item[o] ideal optimal solution using perfect ADN model.
    \end{tablenotes}
  \end{threeparttable}
\end{table}

\Cref{fig:delayed,fig:delayed141} shows the results in online application.
With OLDC, MACSAC has achieved smaller active power loss and VVR than CSAC in this scenario.
The obviously better performance justifies multi-agent RL especially MACSAC with OLDC as an outstanding solution for VVC in ADNs.

Comparing MACSAC and MADDPG, though both algorithms are conducted under OLDC, MACSAC converges to much better power loss and VVR with more stable performance.
Besides, though MACSAC cannot converge to the exact optimal solution as a oracle method, it outperforms other RL-based methods apparently in the online application.
Such significant privilege over MADDPG in terms of active power loss and VVR supports MACSAC as a preferred multi-agent RL algorithm for VVC in ADNs. With the improved voltage performance and smaller voltage violations, the voltage profile can be kept  in an acceptable range by a designed voltage range $[\underline{V},\overline{V}]$.

\subsection{Unbalanced Active Distribution Network}
\label{sub:3phase}
Since the proposed method does not require the inner detailed model of the controlled network, it is also applicable for the unbalanced ADNs with slight modifications on the observation spaces and reward functions. The three phase voltage magnitudes of $j$th node are defined as $V^a_j, V^b_j, V^c_j$. In the observation $o_i=(\mathbf{P}_i,\mathbf{Q}_i,\mathbf{V}_i,\mathbf{P}^e_i,\mathbf{Q}^e_i)$, $\mathbf{V}_i$ is updated from single-phase voltage magnitudes $[V_{j\in\mathcal{N}_i}]$ to three-phase voltage magnitudes $[V^a_{j\in\mathcal{N}_i},V^b_{j\in\mathcal{N}_i},V^c_{j\in\mathcal{N}_i}]$.
As for the reward function, the voltage violation rate penalty $R_V(\mathcal{N}_i,t)$ is updated as $-\sum_{j\in\mathcal{N}_i}\left[[\bar V_j(t)-\overline{V}]_{+}^2 + [\underline{V}-\bar V_j(t)]_{+}^2\right]$, where $\bar V_j = \sum_{h\in\{a,b,c\}} V^h_j / 3$. In the proposed method, only the input and output data are utilized to learn and control, so the algorithm part still holds.
In order to test the validity of the proposed method on unbalanced ADNs, the IEEE 37-bus test feeder is installed with three PV inverters and one SVC and assumed as 3 stations. The simulation is carried out via OpenDSS. Detailed parameters are given in the supplemental file \cite{zzsupple}.

\begin{figure}[htbp]
  \centering
  \includegraphics[width=0.95\linewidth]{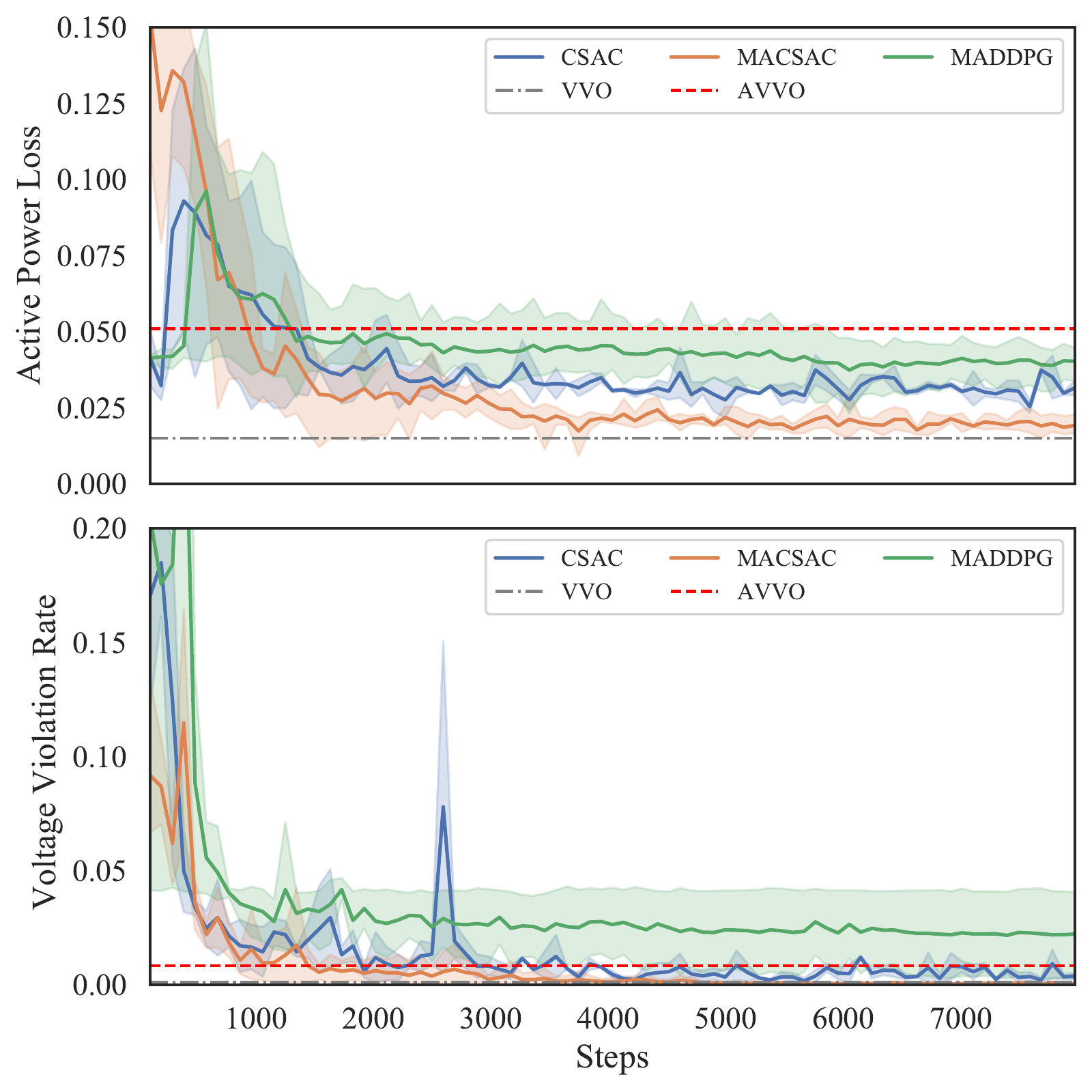}
  \caption{Online application performance with real-world simulation of three-phase unbalanced IEEE 37-bus test feeder.}
  \label{fig:3phases}
\end{figure}

The experiment is carried out with the same online application setting as \cref{sub:online}. The learning process and control performance is illustrated in \cref{fig:3phases}.

In \cref{fig:3phases}, it can be observed that the proposed method is still applicable in the unbalanced distribution network since MACSAC has achieved similar performance to the optimal point. Also, comparing with the centralized CSAC method, the proposed MACSAC with OLDC has achieved smaller active power loss and VVR. Comparing with MADDPG, the proposed method converges to better operation point with more stable performance. Overall, the validity and privilege of the proposed method in the unbalanced distribution networks are supported with this case.

\subsection{Analytical Comparison}
In order to show the novelty and improvement of the proposed method comparing with existing works, \cref{tbl:comparison} has been summarized.
Comparing with three typical baselines including AVVO (optimization method with approximate models), CSAC \cite{8944292} and MADDPG \cite{NIPS2017_7217}, the proposed MACSAC with OLDC has its unique features in several aspects according to the numerical experiments.
Firstly, together with CSAC and MADDPG, the proposed MACSAC with OLDC learns the VVC strategy in a model-free manner, and can achieve near optimal performance without accurate models.
Secondly, MACSAC with OLDC as well as MADDPG is designed for decentralized control and centralized learning. It keeps the control policies in the decentralized controllers and can realize fast control with local measurements and is robust to communications. 
Finally, MACSAC with OLDC has introduced stochastic policies, maximum entropy regularization, explicit voltage constraints, detailed timing design and adjustable sampling ratios comparing with existed MARL algorithms. These targeted methods have significantly improved the control performance in terms of reducing power losses and mitigating voltage violations as well as the convergence and stability of learning.

\begin{table*}[h]
  \centering
  \caption{Multi-dimensional Comparison of the Proposed Method With Baseline Methods}
  \begin{tabularx}{\linewidth}{lXXXXX}
   \toprule
    Methods & Control Structure & Feature of Methodology & Robustness and Convergence & Performance\\
    \midrule
    AVVO & 
      Centralized optimization & 
      1. Model-based; 2. Bad performance with inaccurate models; 3. No exploration needed. &
      It does not need learning process but fails when the communication is lost. &
      With approximate models, there could be severe voltage violations and high line losses. \\
    & & & & \\
    CSAC &
      Centralized learning and centralized control &
      1. Model-free; 2. Single-agent 3. Stochastic policy with soft constraints; 4. Voltage constraints are explicitly considered. &
      With all samples collected to the centralized agent, it converges better than DDPG; but when the communication is lost, the control is interrupted and fails. &
      It can effectively optimize the line losses and voltage violations. However, if the communication delays are considered, the agent is less efficient.\\
    & & & & \\
    MADDPG &
      Centralized learning and decentralized control &
      1. Model-free; 2. Multi-agent; 3. Deterministic policies with random exploration; 4. Voltage violations are penalized with fixed parameters. &
      The control process is carried out locally, and is robust to the communication; however, the convergence can be relatively slow with more samples needed. &
      With sufficient training samples and the communication delays simulated, the agent can achieve similar performance as the centralized RL method CSAC. \\
    & & & & \\
    \textbf{MACSAC with OLDC} &
      Centralized learning and \textbf{decentralized control} with \textbf{detailed timing design and adjustable sampling ratio} &
      1. Model-free; 2.\textbf{Multi-agent;} 3. \textbf{Stochastic policy} with soft constraints; 4. \textbf{Voltage constraints are explicitly considered.} &
      The control process is carried out locally, and is robust to the communication; the \textbf{convergence is fast and stable} with stochastic policies and soft constraints. &
      Considering the conmmunication delays, the decentralized agents can \textbf{achieve better performance than CSAC since the local controllers are much faster.} \\
    \bottomrule
  \end{tabularx}
  \label{tbl:comparison}
\end{table*}

\section{Conclusion}
\label{sec:conclusion}
An online multi-agent RL framework OLDC and the corresponding algorithm MACSAC are proposed for VVC to optimize the reactive power distribution in ADNs without the knowledge of accurate model parameters.
With the consideration of distributed stations with high speed IB-ERs in ADNs, the online multi-agent learning and decentralized control framework can both learn the control experiences continuously to meet the incomplete model challenge, and make decision locally to keep high control speed.
Instead of the existing MADDPG, we propose the safe and efficient MACSAC with maximum entropy regularized stochastic policies and explicitly modelled constraints, which prevents optimization failure and ameliorates training robustness.
Numerical studies on ADNs represented by the modified 33-bus and 141-bus test cases indicate that the proposed MACSAC outperforms the benchmark methods in the online application.
Also, it is demonstrated that OLDC has remarkable superiority for online multi-agent RL-based VVC with extraordinary efficiency and robustness to various computing and communication conditions.

In the future work, transfer learning or meta learning with approximate models or historical data can be studied to provide the proposed MARL-based method with a soft start.
The application of the proposed OLDC to other distributed or decentralized control problems is also a promising research direction. With improved performance, MACSAC has the potential to handle more complex control problems. 

\appendices
\section{Hyperparameters}
The hyperparameters of MACSAC, MADDPG and CSAC used in this paper are shown in \cref{tbl:params}. If the parameters is different in the 33-bus, 141-bus and IEEE 37-bus cases, they would be listed in $\{\cdot,\cdot,\cdot\}$.
\begin{table}[h]
  \centering
  \caption{Algorithm Hyperparameters}
  \label{tbl:params}
  \begin{tabular}{@{}lll@{}}
  \toprule
  Algo. & Parameter & Value \\ \midrule
  Shared & optimizer & Adam \\
   & non-linearity & ReLU \\
   & replay buffer size & $4\times 10^{5}$ \\
   & no. hidden layers & $\{2, 3, 3\}$ \\
   & size of hidden layers & $256$ \\
   & episode size & $96$ \\
   & $\eta$ & $0.995$ \\
   & $\lambda$ & $10^{-3}$ \\ \midrule
  CSAC   & $\alpha$ & $\{0.1, 0.3, 0.13\}$  \\ 
         & $\overline{J}^c$ & $0$ \\
         & learning rate $\sigma$ & $1e-3$ \\ \midrule
  MACSAC & $\alpha_i$ & $\{[0.1]_{1\dots 4}, [0.21]_{1\dots 5}, [0.13]_{1\dots 3}\}$  \\
         & $\overline{J}^c_i$ & $0, \forall i$ \\
         & learning rate $\sigma_i$ & $\{1e-3, \forall i\}$ \\ \midrule
  MADDPG & noise & $\{0.07, 0.05, 0.10\}$ \\
         & learning rate $\sigma_i$ & $\{1e-3, \forall i\}$ \\ \bottomrule
  \end{tabular}
\end{table}

\ifCLASSOPTIONcaptionsoff
  \newpage
\fi

\bibliographystyle{IEEEtran}
\bibliography{paper}


\end{document}